\documentclass{ws-procs975x65}

\begin{document}

\title{Another coincidence problem for $\Lambda$CDM?}

\author{\uppercase{Pim van Oirschot}$^*$}

\address{Department of Astrophysics/IMAPP, Radboud University Nijmegen,\\
P.O. Box 9010, 6500 GL Nijmegen, The Netherlands\\
$^*$E-mail: P.vanOirschot@astro.ru.nl}

\author{\uppercase{Juliana Kwan}} 

\address{High Energy Physics Division, Argonne National Laboratory, Lemont, IL 60439, USA}

\author{\uppercase{Geraint F. Lewis}}

\address{Sydney Institute for Astronomy, School of Physics, A28, \\
The University of Sydney, NSW 2006, Australia}

\begin{abstract}
Over the last nine years of cosmic microwave background observations, 
the Wilkinson Microwave Anisotropy Probe ($WMAP$)
results were consistent with a $\Lambda$CDM cosmological model in which the age of the Universe is
one Hubble time, and the time-averaged value of the
deceleration parameter is consistent with zero. 
This curious observation has been put forward as a new coincidence problem for the 
$\Lambda$CDM concordance cosmology, which is in fact a `greater' coincidence than the
near equality of the density parameters of matter and the cosmological constant.
At the moment of writing these conference proceedings, the Planck Collaboration has released
its first cosmological data, which revealed a small shift in the $\Lambda$CDM 
cosmological parameters when compared to $WMAP$. 
We show that under the assumption of a spatially flat $\Lambda$CDM cosmology, 
Planck's results remove this coincidence problem for $\Lambda$CDM at greater than  
99\% confidence level.
\end{abstract}


\bodymatter

\vspace{0.2cm}
\noindent Komatsu et al.\cite{Komatsu:2011} presented the 
combination of seven-year data from $WMAP$ and improved astrophysical data,
yielding constraints on the basic six parameters of a flat $\Lambda$CDM model.
Their maximum likelihood parameter for $\Omega_{\Lambda_o}$ is 0.728, and
$\Omega_{\Lambda_o} = 0.725 \pm 0.016$ is their mean 
and 1$\sigma$ of the posterior distribution\footnote{See 
the columns `$WMAP$+BAO+$H_o$ ML' and `$WMAP$+BAO+$H_o$ Mean' in their Table 1. 
Throughout this text, a subscript $o$ denotes the present day value.}.
We now show that precisely this value of $\Omega_{\Lambda_o}$ is related to
a present day age of the Universe that is consistent with one 
Hubble time if we assume a spatially flat $\Lambda$CDM cosmology, 
and that the coincidence is significantly weakened in the Planck data.

Starting from Einsteins equations of General Relativity and the Friedmann-Lema\^itre-Robertson-Walker metric, one can write the Friedmann equation for a spatially flat Universe as
\begin{equation}
\frac{\mathrm{d}a}{\mathrm{d}t} = H_o \sqrt{ a^{-1}\Omega_{m_o}  + a^2\Omega_{\Lambda_o}} \label{1}.
\end{equation}
with $a$ the scale factor, $\Omega_{m_o}$ is the present day normalised matter density, $\Omega_{\Lambda_o}$
is the corresponding energy density of dark energy, and $H_o$ the present day value of the
Hubble parameter. 
Note that the equation of state of dark
energy required for the expansion given by Equation~\ref{1} is $w = -1$, equivalent to a standard 
cosmological constant.
Furthermore, the energy density in radiation has been neglected, as it was only dominant in the early 
stages of the Universe.
Integration of this equation results in an expression for the age of the Universe in terms of $a$:
\begin{equation}
H_o t = \frac{2}{3\sqrt{\Omega_{\Lambda_o}}} \cosh^{-1}\left(\sqrt{1+ \frac{a^3\Omega_{\Lambda_o}}{\Omega_{m_o}}}\right). \label{2}
\end{equation}
Today, $t = t_o$ and $a_o \equiv 1$,
which means that $H_o t_o = 1$, if and only if
\begin{equation}
e^{-3 \sqrt{\Omega_{\Lambda_o}}} (\Omega_{\Lambda_o}-1)+e^{3 \sqrt{\Omega_{\Lambda_o}}} (\Omega_{\Lambda_o}-1)+2 \Omega_{\Lambda_o}+2 = 0
\label{x}
\end{equation}
where, through flatness, it is assumed that the density parameter of matter $\Omega_m = 1-\Omega_\Lambda$.
Because equation~\ref{2} requires $\Omega_{\Lambda_o} > 0$,
the only solution to equation~\ref{x} is $\Omega_{\Lambda_o} \approx 0.737125$.
This special value for $\Omega_{\Lambda_o}$ laid within the 1$\sigma$ of the posterior distribution
after seven-year $WMAP$+BAO+$H_o$ observations quoted above\cite{Komatsu:2011}, however
in the nine-year $WMAP$+BAO+$H_o$ 
cosmology results $\Omega_{\Lambda_o} = 0.712 \pm 0.010$\cite{Hinshaw:2012}.
In general, we can write the time-averaged deceleration parameter $\langle q \rangle$ 
as a function of $a$:
\begin{equation}
\langle q \rangle + 1 = \frac{1}{tH} =\left[\frac{2}{3}\sqrt{\left(\frac{\Omega_{m_o}}{a^3 \Omega_{\Lambda_o}} + 1\right)} \cosh^{-1}\left(\sqrt{1+ \frac{a^3\Omega_{\Lambda_o}}{\Omega_{m_o}}}\right)\right]^{-1} \label{5}
\end{equation}
In the top right panel of Figure~1, we plot $\langle q \rangle$ using the 95\% confidence levels
of $\Omega_{\Lambda_o}$ from both the Planck data \cite{Planck:2013} ($0.686^{+0.037}_{-0.040}$)
and the WMAP 9-year data \cite{Hinshaw:2012} ($0.721 \pm 0.050$) without
BAO+$H_o$ priors. 
The probability density function of the posterior distribution is indicated by the colorscale.
Note that $\langle q \rangle_o$ is close to zero only during a brief period
in cosmic time, and with the small errors on the derived value of $\Omega_{\Lambda_o}$
it is remarkable that $\langle q \rangle_o \approx 0$ in the WMAP 9-year data\cite{Hinshaw:2012}.
However, $\langle q \rangle_o \ne 0$ for the Planck data\cite{Planck:2013}, which is clearest in
the zoom-in in the top right panel of Figure~1 near $a_o$.
This is an updated version of the top left panel in Figure~1, in which the 
Komatsu et al. $WMAP$+BAO+$H_o$ data\cite{Komatsu:2011} is used.

The coincidence that $\langle q \rangle_o \approx 0$ was first discovered by Lima\cite{Lima:2007}.
Figure~1 is an adapted version of Figure~1 in van Oirschot et al.\cite{van-Oirschot:2010},
who also showed that $\langle q \rangle_o \approx 0$ is the same as the $R_h \approx ct$ 
coincidence reported by Melia et al.\cite{Melia:2007}.
Comparing this coincidence with the well known coincidence problem for $\Lambda$CDM,
the near equality of the density parameters of matter and the cosmological constant,
we see that this new coincidence is in fact a `greater' one (see the bottom panels of
Figure~1). Whereas the density parameters were equal already a few Gyrs ago,
$\langle q \rangle = 0$ happens instantaneously, where the red and blue solid line
(bottom left panel) or red and green line (bottom right panel) cross.

Recently, Melia et al. proposed the ``$R_h = ct$" universe to naturally explain the coincidence
and touted that it is superior to $\Lambda$CDM in explaining our observations of the Universe
\cite{Melia:2012,Melia:2012a,Melia:2013}.
However, see Bilicki and Seikel\cite{Bilicki:2012} and Lewis\cite{Lewis:2013a} for arguments against 
the $R_h = ct$ universe.
The Planck Collaboration \cite{Planck:2013} finds $\Omega_{\Lambda_o} = 0.686^{+0.048}_{-0.052}$ 
with 99\% confidence, which indicates that we have to go
beyond the 99\% confidence level to arrive at $R_h=ct$.
Using the mean and 1$\sigma$ of the $\Omega_{\Lambda_o}$ posterior distribution from Planck 
gives $R_{h_o} = 1.05 \pm 0.02 \ ct_o$, or $\langle q \rangle_o = 0.05 \pm 0.02$.
Therefore, we can exclude the proposed $R_h = ct$ universe on the basis of the Planck data 
and the new coincidence problem for $\Lambda$CDM is significantly weakened.

\begin{figure}[h]
\begin{center}
\psfig{file=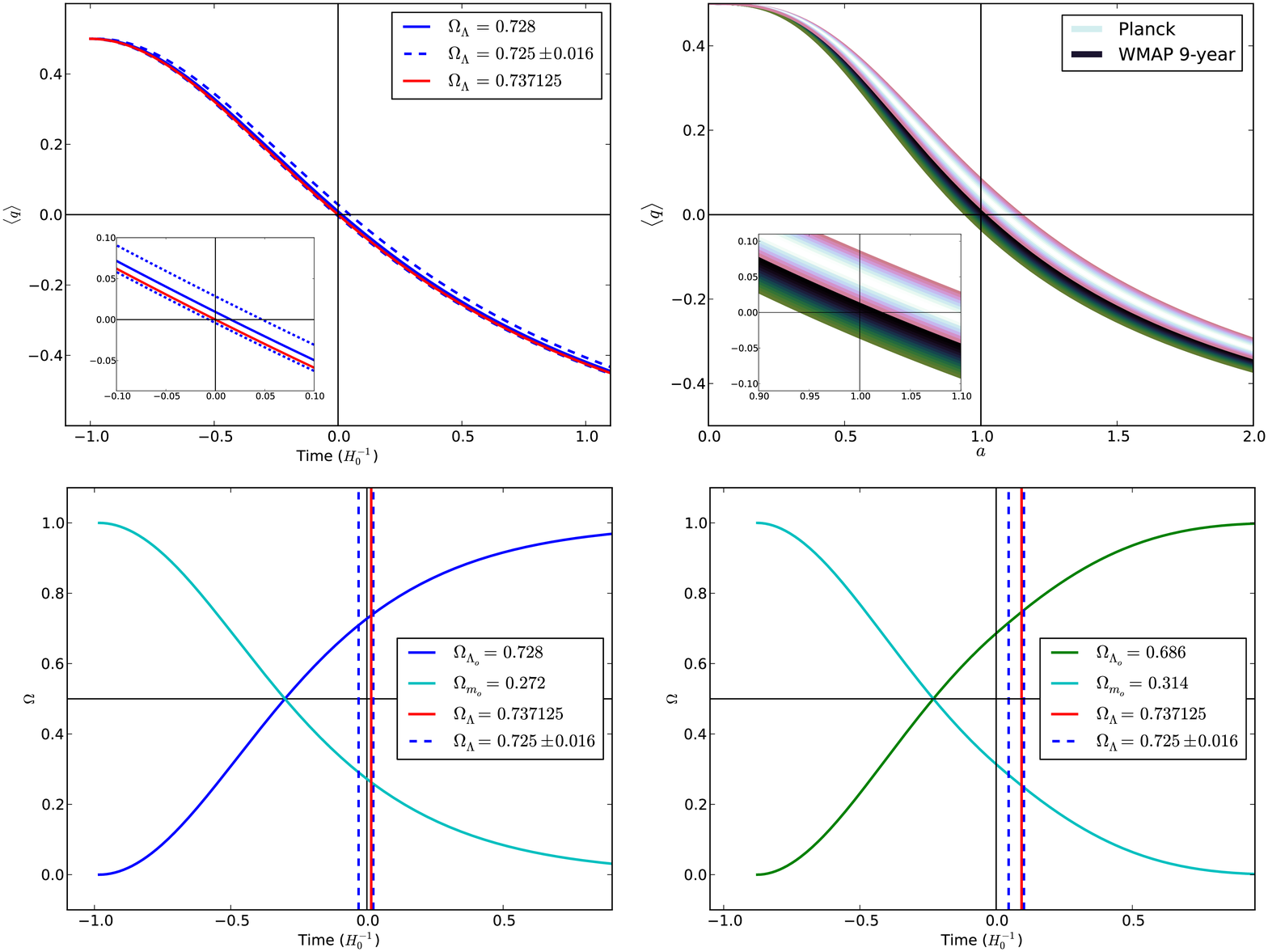,width=0.99\textwidth}
\end{center}
\caption{Top left: The blue dashed and solid lines correspond to the 68\% confidence levels 
on the mean value of the posterior distribution and the maximum likelihood 
of $\Omega_{\Lambda_o}$ from Komatsu et al.\cite{Komatsu:2011}. 
The red line shows the cosmological model
in which $\langle q \rangle_o = 0$, thus $\Omega_\Lambda = 0.737125$.
Top right: An updated version of the top left panel, comparing 
$\langle q \rangle$ as a function of the cosmological scale factor, $a(t)$,
using the 95\% confidence levels of $\Omega_{\Lambda_o}$ in the WMAP 9-year and Planck data.
The brighter (darker) the colors, the more confident the Planck (WMAP 9-year) data.
Bottom panels: The Komatsu et al.\cite{Komatsu:2011} maximum likelihood 
of $\Omega_{\Lambda_o}$ is used for the blue solid line in the left panel, 
that from the Planck Collaboration for the green line in the right panel.
The cyan lines indicate the evolution of the corresponding density parameter of matter.
The coincidence is apparent in the left panel, because 
$t_0$ lays between the dashed lines ($\sim$ 743 Myr) that indicate 
the 1$\sigma$ of the posterior distribution 
of $\Omega_{\Lambda_o}$ from Komatsu et al.\cite{Komatsu:2011},
but it has disappeared in the right panel.}
\label{aba:fig1}
\end{figure}

\bibliographystyle{ws-procs975x65}
\bibliography{main}

\end{document}